\documentclass[aps,pre,twocolumn,showpacs,floatfix]{revtex4}
\usepackage{graphicx,color}
\usepackage{amsmath,amssymb}
\usepackage{hyperref}

\begin{document}
\title{Scaling relation for determining the critical threshold for
continuum percolation of overlapping discs of
two sizes}

\author{Ajit  C. Balram}
\email{cb.ajit@iiserpune.ac.in}
\affiliation{Department of Physics,
Indian Institute of Science Education and Research\\
Sai Trinity Building,
Garware Circle, Sutarwadi, Pashan,
Pune  411 021, India}

\author{Deepak Dhar}
\email{ddhar@theory.tifr.res.in}
\affiliation{Department of Theoretical Physics, Tata Institute of Fundamental Research \\
1 Homi Bhabha Road, Mumbai  400 005, India.}

\date{\today}
\begin{abstract}

We study continuum percolation of  overlapping circular discs of two sizes. We propose a
phenomenological scaling  equation for the increase in the effective size of the larger discs due to the presence of the smaller discs. The critical percolation threshold as a function of the ratio of sizes of discs, for different values of the relative areal densities of two discs, can be described in terms of a scaling function of only one variable. The recent accurate Monte Carlo estimates of critical threshold by Quintanilla and Ziff [Phys. Rev. E, 76 051115 (2007)] are in very good agreement with the proposed  scaling  relation.

\end{abstract}


\maketitle

In recent years, there has been a lot of interest in studying continuum percolation, owing to its many applications. For a review on continuum percolation, see \cite{meester_book}. Continuum percolation of overlapping objects of various sizes and shapes, spheres and discs \cite{scher,consiglio,quintanilla01,gawlinski,phani}, ellipsoids \cite{sastry}, plates \cite{yi}, sticks \cite{balberg}, oriented cubes \cite{stanley} etc., has been studied. In applications like the modeling of porous media, one of the most important parameters is the distance from percolation threshold, and several approximation schemes have been proposed to determine the percolation threshold for different types of disorder.

In this paper, we discuss the case of continuum percolation of  overlapping discs of two sizes in a plane. We propose a phenomenological  equation for the increase in the effective size of the larger discs in the presence of the smaller discs. We check our theory against data on critical thresholds by Quintanilla and Ziff \cite{quintanilla_ziff}. The agreement is found to be very good.

We consider a percolation model  of a mixture of circular discs of two sizes randomly placed  in a  plane. Consider a finite area $S$ and randomly drop discs in $S$.  The probability that a given small areal element $dA$ contains the center of a dropped disc is $ndA$, independent of other discs. Once a center of the disc is chosen, it is assigned a radius $R_1$ with probability $f$, and $R_2$ with probability $(1 - f)$. We denote the ratio of radii $R_1 / R_2 $ by $\lambda$. The number density of discs with radius $R_1$ is then $n_1 = f n$, and that of  radius $R_2$ is $n_2 = ( 1 -f) n$. The  total number density of discs, irrespective of radius, is $n =n_1 + n_2$. We propose an approximate formula for the critical percolation threshold in terms of $\lambda$ and $f$.  We express this function of two variables in terms of the function $\xi(A)$ which gives the correlation length $\xi$ as  a function of the areal density $A$ of single-sized discs.

The earliest proposal for determining the critical threshold for overlapping discs was by Scher and Zallen \cite{scher}. They noted that the total covered fractional area at critical threshold was nearly constant for a  mixture of discs of different sizs, if the polydispersity of the mixture was small.  However, if the polydispersity is large, and one takes  discs   with several  different radii,  the total covered fraction at critical threshold can be made as close to one as we wish \cite{phani}.  The original heuristic arguments have been made rigorous later \cite{meester}.

We start by summarizing the qualitative arguments of \cite{phani}. Let us assume,  without  any loss of generality, that the $R_1 < R_2$. We consider the plane on which   the smaller discs of radius $R_1$ each have been thrown in randomly with  $n_1$ discs per unit area. The areal density of these discs is then $A_1 = \pi R_1^2 n_1$. Note that  $A_1$ is a dimensionless number giving the ratio of total  area of discs thrown in to the area of the plane.  In the
case of percolation of discs of equal radii, the areal density of the discs at the percolation threshold is independent of the size of the discs. Let this critical value of $A$ be denoted by $A^*$. We assume that $A_1$ is below  critical threshold $A^*$, and the small discs by themselves do not percolate.  From numerical simulations, the value of $A^* $ is known quite accurately $A^*  \approx 1.128085$.  The corresponding value of the covered area fraction is given by $ \phi^* = 1 - \exp( - A^*) \approx 0.6763475(5)$\cite{quintanilla_ziff}.

The two point correlation function, $G(r)$, is defined as as the probability that two points at a distance $r$ from each other, chosen at random, belong to the same cluster when only the smaller discs have been dropped.  Below criticality, this decays exponentially with distance, i.e., $G(r) \sim \exp( -r/\xi_1)$.  And using simple scaling invariance of the problem $R_1 \rightarrow \alpha R_1$, we have
\begin{equation}
\xi_1( A_1) = R_1  g( A_1)
\end{equation}
where the function $g(x)$  determines how the correlation length varies with areal density, and is  independent of $R_1$.

Now we throw in a single disc of the larger radius $R_2$, and look at the cluster of discs that are connected to this
single large disc. Then, each such cluster looks like a somewhat bigger fuzzy disc of size $R_2 + \Delta R_2$. Let us assume that the variation between different clusters
may be neglected. This approximation is quite good if $R_2  \gg \xi_1$, but less valid if $R_2/\xi_1 $ is not so large. The percolation problem can then be considered as a percolation of these larger effective discs. The number density $n_2^*$ of these effective larger equal-sized discs of radius $R_2 + \Delta R_2$ that have to be dropped to reach criticality is given by
\begin{equation}
n_2^{*}\pi ( R_2 + \Delta R_2)^2 = A^*
\end{equation}
We will consider this equation as the {\it definition of} $\Delta R_2$.

In \cite{phani}, the simple approximation
\begin{equation}
\Delta R_2 \approx c~\xi_1
\label{eq1}
\end{equation}
was used, where $c$ is some constant of order one. This  gives the correct limiting behavior that for any initial density $A_1$ of the smaller discs, the critical value of the areal density of larger discs  $A_2^*(R_2) $ tends to $A^*$ as  $R_2$ tends to infinity, keeping $A_1$ fixed. Also,  the other  limit when  we keep $A_2$ fixed at any value below $A^*$,  and slowly increase $A_1$ till we reach critical percolation, then the critical value of $A_1^*(R_1)$  to reach criticality tends to $A^*$ as $R_1$ tends to zero \cite{meester}.

However, Eq.(\ref{eq1}) strongly underestimates the
value of $\Delta R_2$. Consider two discs of radius $R_2$ thrown in a sea of randomly dropped smaller discs of areal density $A_1$. Call these discs 1 and 2 and, let the minimum distance between these discs be denoted by $D$ (Fig. 1). We denote  by $Prob_D (1\leadsto 2)$ the probability that there is a path of overlapping smaller discs between the  larger discs, and they belong to the same cluster.  Thus, $Prob_D ( 1\leadsto 2)$ is a measure of the connectivity correlations in the problem of percolation of single-sized discs.

Clearly,  $Prob_D (1\leadsto 2)$ is a decreasing function of the separation $D$, which will decrease exponentially from $1$ to $0$, as $D$ varies from $0$ to infinity. For large $D$,  this decreases as $\exp( -D/\xi_1)$. The dependence of this  on $R_2$ comes from the fact that the prefactor of the exponential would depend on $R_2$. Also, for $D$ comparable to $\xi_1$, the $D$-dependence can not be approximated well by a simple exponential. However, we can define an effective size $\Delta R_2^{\mathrm {eff}}$ by the requirement that this probability is a fixed value, say $1/2$, when $D = 2 \Delta R_2^{\mathrm {eff}}$. Then, a better estimate of $\Delta R_2$ than Eq. (\ref{eq1}) is given by
\begin{equation}
 \Delta R_2 \approx \Delta R_2^{\mathrm {eff}}.
\end{equation}

\begin{figure}
\centering
\includegraphics[width=3.0in]{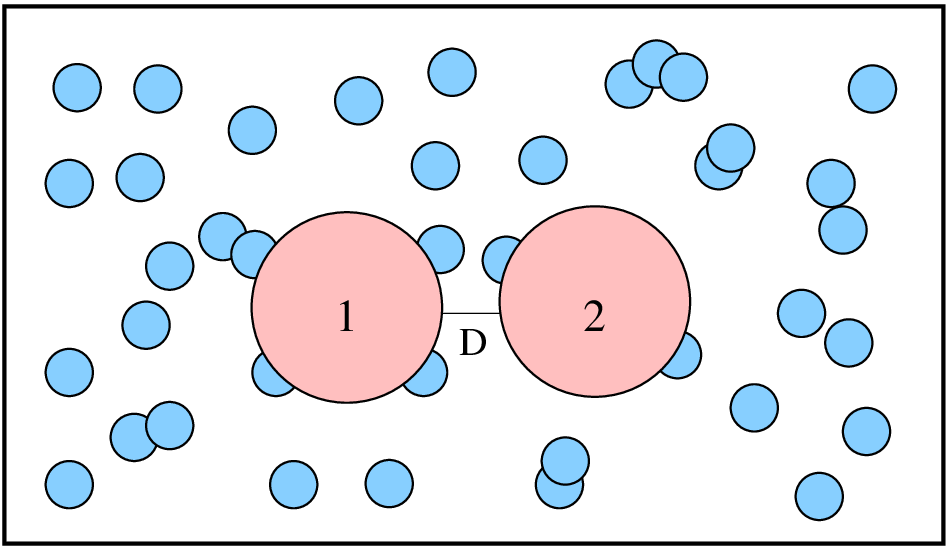}
\caption{Two large discs of radius $R_2$ in a background of randomly dropped smaller discs. The least separation between the discs is $D$.}
\label{fig:1}
\end{figure}

The $\Delta R_2^{\mathrm {eff}}$ as defined is a function of $ R_1, R_2 $ and $ n_1$ (or, equivalently $\xi_1$). For $D$ comparable to $\xi_1$, we cannot use the large $D$ exponential decay of  $Prob_D (1\leadsto 2)$  to estimate $\Delta R_2^{\mathrm {eff}}$. However, if $R_1 \ll  \xi_1$, then we can assume that the leading dependence is from $\xi_1$, and correction terms involving powers of $R_1/\xi_1$ can be neglected.  Then, $\Delta R_2^{\mathrm {eff}}$, to leading order,  is only a function  of $R_2$ and $\xi_1$. Using the fact that the probabilities are invariant if all distances are scaled by same factor, we get

\begin{equation}
\Delta R_2^{\mathrm {eff}} = \xi_1  h( R_2/\xi_1)
\end{equation}
where $h(x)$ is some, as yet unspecified,  scaling function of its argument $x$. Now, clearly, $Prob_D (1\leadsto 2)$ is a monotonically increasing function of $R_2$, which tends to $1$ as $R_2$ tends to infinity, keeping $D$ fixed, as then the problem is that of percolation in a very  long strip, and somewhere or other, there will be a connection of smaller discs. This implies that  $\Delta R_2^{\mathrm {eff}}$ must tend to infinity if $R_2$ tends to infinity. Also, in the case $R_1 \ll R_2 \ll \xi_1$,  it must tend to infinity as $\xi_1$ tends to infinity.  The simplest form of $h(x)$ that is consistent with these requirements is a simple power-law form, which gives
\begin{equation}
\Delta R_2^{\mathrm {eff}} = k ~\xi_1^{a} ~R_2^{1 - a}.
\label{eq2}
\end{equation}

Here $k$ is some constant of order $1$. The main improvement in this form over Eq.(\ref{eq1}) is the inclusion of dependence on $R_2$.

The power-law dependence of  $R_2^{\mathrm {eff}}$ on $R_2$ is seen  most easily by considering a perturbation expansion of $Prob_D (1\leadsto 2)$ in powers of $n_1$. Let
\begin{equation}
Prob_D (1\leadsto 2) = n_1 F_1( D, R_2) + n_1^2 F_2(D,R_2) + \ldots
\end{equation}

\begin{figure}
\centering
\includegraphics[width=3.5in]{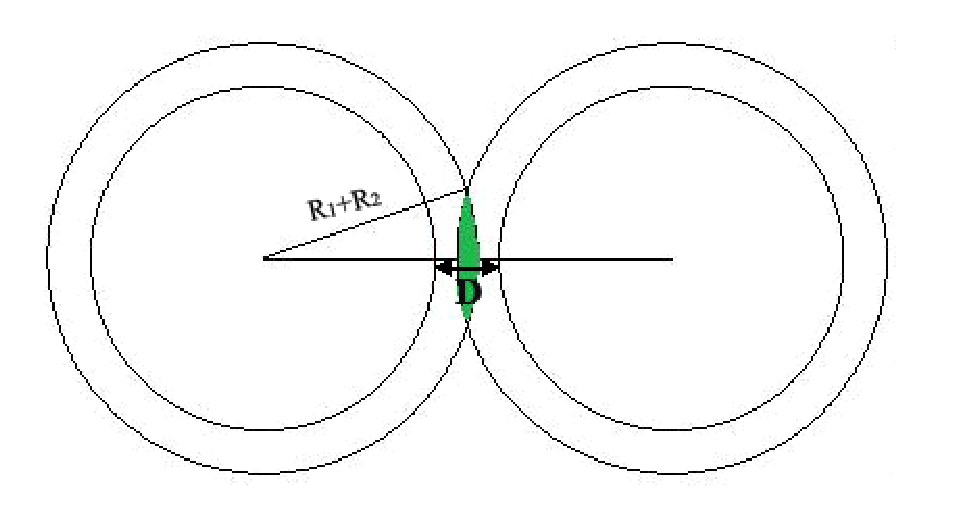}
\caption{Two large discs of radius $R_2$ with  separation between the discs $D$.  A larger circle of radius $R_1 + R_2$ is drawn surrounding each of the discs. If the center of any small disc falls in the  intersection region  (shown shaded) of the larger circles, it forms a connecting path betwen them.}
\label{twop}
\end{figure}

In the first order in $n_1$, the configurations that contribute to $Prob_D (1\leadsto 2)$ are those
where a single small disc overlaps with both the bigger discs. This is possible only if $D < 2 R_1$,
and in that case, if there is at least one small disc in the region which is within a distance $R_1 + R_2$  from the centers of the discs $1$ and $2$ ( See Fig. \ref{twop}). For small $n_1$, the probability of this event is proportional to the area of the shaded region in Fig. \ref{twop}. Using elementary geometry, it is easily seen that  for $R_2 \gg R_1$, the area is proportional to $R_2^{1/2} ( 2 R_1 - D)^{3/2}$.
Thus we get
\begin{equation}
F_1( D, R_2) \sim  R_2^{1/2} ( 2 R_1 - D)^{3/2}, {\rm ~~ for ~~} 0 \leq D \leq 2 R_1.
\end{equation}
Thus, we see that $F_1$, and by extension $Prob_D( 1\leadsto 2)$ has  a strong dependence on $R_2$.
Of course, for $\xi_1 \gg R_1$, higher order terms in $n_1$ make significant contribution, and they
would change the precise form of the functional dependence on $R_2$.

Again, we assume that the larger discs act as discs of radius $R_2 + \Delta R_2$, with $\Delta R_2 \simeq \Delta R_2^{\mathrm {eff}}$, given by Eq.(\ref{eq2}).
Expressing $n_2$ in terms of $A_2$, the areal density of the larger discs, the criticality condition may be written as
\begin{equation}
\frac {\Delta R_2}{R_2} = {\sqrt{ A^*/A_2} - 1} \approx k [\lambda g(A_1)]^{a}
\label{eq3}
\end{equation}

The above equation is clearly invariant under scaling of all lengths by the same factor. We can determine the value of $a$, in the limit $\xi_1 \gg R_2$. Then, assume $A_1 = A^*( 1 - \epsilon)$. Then, $\epsilon \ll 1$ implies that $\xi_1 \gg R_1$.

Clearly, the number density of additional discs of radius $R_2$ required to reach criticality would be less than with discs of size $R_1$. Hence, in terms of areal densities, this bound becomes $A_2^* < A^* \epsilon \lambda^{-2}$.
Also, as discussed in \cite{phani}, the total areal density of discs at criticality is greater than $A^*$ when all discs are not of same size,   $ A_2^* \geq \epsilon
A^*$,  Thus, $A_2^* \sim \epsilon$.  Then, $\Delta R_2 \sim \epsilon^{-1/2}$. Since it is known that  $g(x) \sim ( A^{*}-x)^{-\nu}$ for $x$ near $A^{*}$, with $\nu = 4/3$. Thus, comparing powers of $\epsilon$ we see that $a = 3/8$.

Our proposed approximation can be directly checked against numerical data. Quintanilla and Ziff have given a very extensive table of data giving different values of $A_1, A_2$  for different values of $R_1/ R_2$, that define critical  surface \cite{data}.
Using Eq.(\ref{eq3}), if we plot $Y = \lambda^{-a} [\sqrt{ A^*/A_2} - 1] $ versus $X = A^* - A_1$, all points should fall on  a single curve $ Y = g(A^* - X)^{a}$.  The result is shown in Fig. 3, where we have plotted data corresponding to five different values of $\lambda= 0.10, 0.20, 0.30$ and $0.50 $  We get a very good collapse.  We do not show other values, in order not to clutter up the figure, but have checked that the collapse is as good with them as well. Note that no free parameters have been used to generate the scaling collapse.

Define $\Phi(x)=[g(x)]^a$. The function $\Phi(x)$, which gives the equation of the curve is in principle calculable if we can solve the problem of percolation probability with single sized discs. As of now, we only know the behavior of $\Phi$ in certain regimes. For small $x$, $\Phi(x) \sim  x$ and for $x$ near $A^{*}$, $\Phi(x)$ varies as $( 1 - x/A^*)^{-1/2}$. Hence we parameterize the curve as
\begin{equation}
\Phi(x) \sim  kx ( 1+ c x) (1 -x/ A^*)^{-1/2}.
\end{equation}
The values $k=0.25$ and $c =2.20$, give a fairly good fit. The  curve using these fitting parameters is also shown in Fig. 2.

\begin{figure}
\centering
\includegraphics[width=3.0in]{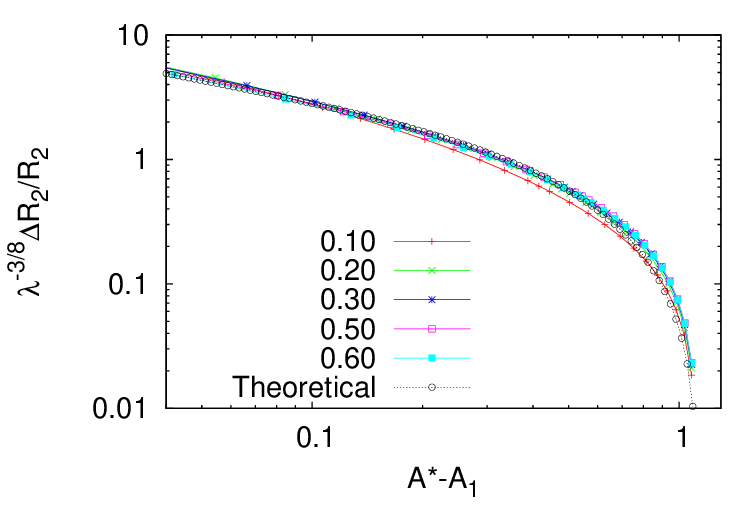}

   \caption{Scaling collapse of the Monte Carlo data of \cite{quintanilla_ziff}. Re-scaled data of $\Delta R_2$ is plotted vs $A^* - A_1$, the deficit in the areal density $A_1$ of smaller discs from the critical value for mono-disperse discs $A^*$, for different values of the ratios of radii $\lambda$.}
\label{fig:2}
\end{figure}

We would like to thank Prof. R. M. Ziff,  for useful correspondence.
ACB thanks Kishore Vaigyanik Protsahan Yojana  for financial support, and TIFR for  hospitality   during his visit. DD would like to acknowledge the financial support from  the Department of Science and Technology, Government of India, through  a JC Bose Fellowship.

\end{document}